\documentstyle[epsf]{article}
\def\boxit#1{\kern4pt\vbox{\hrule\hbox{\vrule\kern8pt\vbox{\kern8pt#1\kern8pt}\ern8pt\vrule}\hrule}}

\def\and{ {\rm and} }

\def\half{ \frac{1}{2} }

\def\half{\frac{1}{2}}
\def\be{\begin{equation}}
\def\ba{\begin{eqnarray}}
\def\ea{\end{eqnarray}}   
\def\gr{\nabla}
\def\ee{\end{equation}}
\def\to{\rightarrow}
\def\tmu{$TH\epsilon\mu\;$}
\def\pd{\partial}
\def\nn{\not\!}
\def\sp{\overline\psi}

\def\al{\alpha}

\def\nv{\hat n}

\begin{document}
\onecolumn
\vfill
\begin{center}
{\Large \bf 
Test of the Equivalence Principle
using Atomic Vacuum Energy Shifts} \\ \vfill
C. Alvarez\footnotemark\footnotetext{email:
calvarez@avatar.uwaterloo.ca} 
and R.B. Mann\footnotemark\footnotetext{email: 
mann@avatar.uwaterloo.ca}\\
\vspace{2cm}
Dept. of Physics,
University of Waterloo
Waterloo, ONT N2L 3G1, Canada\\
\vspace{2cm}
\vspace{2cm}
\today\\
\end{center}
\vfill

\begin{abstract}
We consider possible tests of the Einstein Equivalence Principle
for quantum-mechanical vacuum energies by 
evaluating the Lamb shift transition in a class of non-metric theories of
gravity described by the \tmu formalism. We compute to lowest order
the associated red shift and 
time dilation parameters, and discuss how
(high-precision) measurements 
of these quantities could provide  new 
information on the validity of the equivalence principle. 
\end{abstract}

\vfill

\twocolumn

The Einstein Equivalence Principle (EEP) is foundational to our
understanding of gravity. It ensures that a unique operational
geometry can be attributed to spacetime, thereby guaranteeing
that the effects of gravity on
matter are accounted for in a purely geometric fashion.
Metric theories, such as
general relativity and Brans-Dicke Theory realize this principle by
endowing spacetime with a symmetric, second-rank tensor field $g_{\mu\nu}$ that
couples universally to all non-gravitational fields \cite{kll}. 
Non-metric theories do not have this feature:  they violate universality  by
coupling auxiliary gravitational fields directly to matter,
thereby permitting observers performing local
experiments to detect effects due to their position and/or velocity
in an external
gravitational environment.

Empirical limits on such effects are set by gravitational red-shift  and
atomic physics experiments \cite{redshift,PLC}, each of which
compares relative frequencies of transitions between particular energy
levels that are 
sensitive to any possible position or velocity dependence respectively.
Significantly improved levels of precision for such experiments are anticipated
in the next generation of  gravitational experiments;
for example, gravitational 
red shift experiments could be
improved to one part in $10^{9}$  by placing a hydrogen maser 
clock on board Solar Probe, a proposed 
spacecraft (see ref. \cite{will1} and references therein).
The dominant form of energy governing the transitions these experiments probe
is nuclear electrostatic energy, although violations of the EEP due to
other forms of energy (virtually all of which are associated with baryonic
matter) have also been estimated \cite{HW}.
Potential violations of the EEP due to effects such as vacuum energy shifts,
which are peculiarly quantum-mechanical in origin 
({\it i.e.} are due solely to radiative corrections)
are not as well understood.  In this paper we report on the results of
an investigation into the effects that EEP-violating couplings have on
 Lamb-shift transition energies in Hydrogenic atoms. Details will appear
in a forthcoming paper \cite{catlamb}.
A test of the EEP for this form of energy provides
us with a qualitatively new empirical window of the foundations of 
gravitational theory.

We begin with the action appropriate for Quantum Electrodynamics in
a background gravitational field:
\begin{equation}\label{act}
\!\!S\!=\!\int\!\! d^4x\sqrt{-g}\!\left[ \sp(i\nn\gr\!+\!e\nn\! A\!-\!m)\psi
\!-\!\frac{1}{4}F_{\mu\nu}F^{\mu\nu}\right]
\end{equation}
where 
$\gr$ is the covariant derivative, 
$F_{\mu\nu} \equiv A_{\nu,\mu} - A_{\mu,\nu}$ and
${\nn\! A}= e^a_\mu \gamma_a A^\mu$, $e^a_\mu$ being the tetrad associated
with the metric. Using the \tmu formalism \cite{tmu} (which 
encompasses a wide
class of non-metric theories in addition to all metric theories)
the general form of the static, spherically
symmetric metric is $g_{\mu\nu}=$diag$(-T,H,H,H)$ and 
$F_{\mu\nu}F^{\mu\nu} = 2(\epsilon E^2-B^2/\mu)$,
where $\vec E\equiv-\vec\gr A_0-\pd\vec A/\pd t$ and
$\vec B \equiv \vec\gr\times\vec A$. 
$\epsilon$ and $\mu$ are
arbitrary functions of the Newtonian background potential
$U= GM/r$ (which approaches unity as $U\to 0$) as are $T$ and $H$,
which in general will depend upon the species of particles within the
system (electrons in the present case).

Consider an atom that moves with velocity $\vec u$ relative to the
preferred frame.  The spacetime scale of atomic systems 
permits us to ignore spatial
variations of $T$, $H$, $\epsilon$ and $\mu$; using this and a Lorentz
transformation to transform fields and coordinates from the
preferred frame to the rest frame of the atom ({\it i.e.} the
moving frame), it may be shown that (\ref{act}) to $ O(\vec u^2)$ reduces to
\cite{gabriel} 
\begin{eqnarray}\label{3}
S&=&\int d^4x\Big\{ \sp(i\nn\pd+e\nn\! A-m)\psi+J_\mu A^\mu\nonumber\\
&+&\half\Big[E^2-B^2+\xi\Big(\vec u^2 E^2-(\vec u\cdot\vec E)^2\\
&+&(1+\vec u^2)B^2-(\vec u\cdot\vec B)^2
  +2\vec u\cdot(\vec E\times\vec B)\Big)\Big]\Big\}\nonumber.
\end{eqnarray}
$J^\mu$ is the current
associated with some external source (taken here to be a pointlike spinless
nucleus).

The dimensionless parameter $\xi = 1-H_0/T_0\epsilon_0\mu_0 \equiv 1-c^2$ 
measures the degree of EEP violation, 
with ``0'' denoting the functions evaluated at the atomic system's 
center of mass and $c$ being the ratio of the limiting speed of electrons
to the speed of light.  The natural scale for $\xi$ 
is set by the magnitude of $U$, which empirically 
is much smaller than unity \cite{will1}, permitting us to perform a perturbative
analysis in $\xi$ and $\vec{u}$ of the radiative corrections associated with
the action (\ref{3}).

Perturbatively solving the field equations associated with (\ref{3})
for the $A_\mu$ produced by a pointlike nucleus of charge $Ze$ at rest in
the moving frame yields \cite{gabriel}
\begin{eqnarray}
\label{po}
A_0 &=& [1-\frac{\xi}{2}(\vec u^2+(\vec u\cdot\hat n)^2)]\phi
\equiv\phi+\xi\phi'
\nonumber\\
\vec A&=&\frac{\xi}{2}[\vec u+\hat n(\vec u\cdot\hat n)]\phi\equiv\xi\vec A\,'
\end{eqnarray}
where $ \hat n=\vec x/|\vec x|$, $\phi=Ze/4\pi|\vec x|$, and 
$\vec\gr\cdot \vec A=0$. The Dirac Hamiltonian may then be written as
\be\label{eqq}
H=H_0+\xi  H',\qquad H'=-e\phi'+e\vec\alpha\cdot\vec A'
\ee
where $H_0$ 
corresponds to the standard Hamiltonian (with Coulomb potential only),
and  the primed fields are defined in (\ref{po}). Using first-order
perturbation theory on the Dirac equation, 
$H|n\rangle=E_n|n\rangle$, we can solve for 
$|n\rangle=|n\rangle^0+\xi|n\rangle'$ and $E_n=E_n^0+\xi E_n'$,
which yields 
\be\label{ED}
E_{2S_{1/2}}-E_{2P_{1/2}}=\xi\frac{u^2}{6}m(Z\alpha)^4+
O\left((Z\alpha)^6\right)
\ee
and so the 
$2S_{1/2}$--$2P_{1/2}$
degeneracy is lifted before radiative corrections are introduced.   
This nonmetric contribution to the Lamb shift
is isotropic in the 3-velocity $\vec u$ of the moving frame and vanishes
when $\vec{u} = 0$.

To lowest order in QED there are two types of radiative corrections to the
energy levels of an electron bound in an external electromagnetic
potential (shown in Fig. 1): the vacuum polarization ($\Pi$) and self-energy ($\Sigma$),
along with a counterterm ($\delta C$) that subtracts the
analogous processes for a free electron.  
The radiative contributions to a state $|n\rangle$ will therefore be
$\delta E_n=\langle n|\Sigma-\delta C +\Pi |n\rangle$.

\begin{figure}[h]
\centering
\leavevmode
\epsfbox[63 136 375 264] {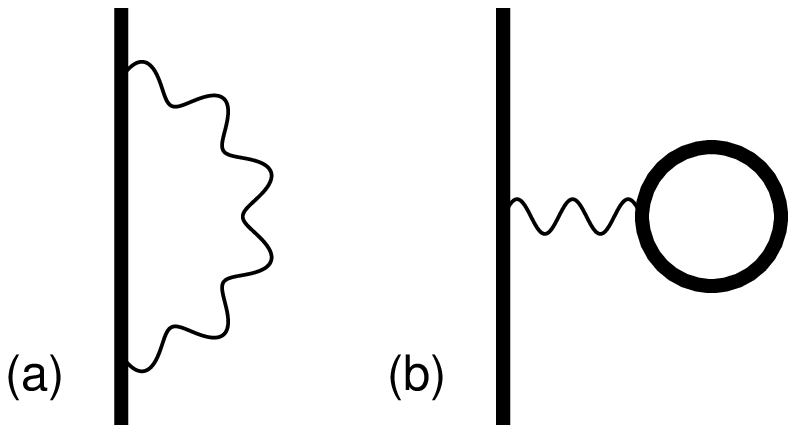}
\end{figure}
{\footnotesize Fig. 1. Radiative corrections of order $\alpha$ : (a) self-energy
 and (b) vacuum polarization.} 

The bold line in Fig. 1 represents the bound electron propagator
$(\nn\! p+e\nn\!\! A-m)^{-1}$, with $p^\mu\equiv(E_n, \vec p)$,
and $A^\mu$ given by (\ref{po}). Upon  adding a gauge fixing term
$[-(1-\xi)(\pd\cdot A)^2-2\xi \pd^0A_0\pd\cdot A]$
to (\ref{act})  we obtain
\begin{equation}\label{11} 
G_{\mu\nu}=-(1+\xi)\frac{\eta_{\mu\nu}}{k^2}
+\xi\frac{\gamma^2}{k^2}\left[\eta_{\mu\nu}\frac{(\beta\cdot k)^2}{k^2}
+\beta_\mu\beta_\nu\right]
\end{equation} 
for the photon propagator, where $\eta_{\mu\nu}$ is the Minkowski tensor with
 signature $(+ - - -)$,  $\beta^\mu\equiv (1,\vec u)$ and
$\gamma^2\equiv(1-\vec u^2)^{-1}$.

As in the metric case, regularization and renormalization processes are needed.
We choose the cutoff method to deal with
the divergences, absorbing them by proper redefinitions of the
parameters of the theory, which now include the \tmu functions. 
Even though we reformulate
quantum electrodynamics without the symmetries given by
local position and local Lorentz invariance, the theory still remains
gauge invariant and the self consistency 
of the Ward Identities is straightforwardly checked.

The Lamb shift calculation involves dealing with bound electron propagators,
which in turn requires an approach adequate
to sort out the external electromagnetic field
dependence. We follow the method given by ref.\cite{BBF} in which the
bound propagator is separated into a term where the external 
potential acts only once, and another term where it acts at least twice. 
After a lengthy calculation similar to that described in ref.
\cite{BBF} we find to the required accuracy
$O(\xi)O(\vec u^2)O(\al(Z\al)^4 )$,
\be\label{Elamb}
\!\!\delta E_n\!\!=\!\!\frac{\alpha}{3\pi m^2}\!\!\left[\! \Big (\!1\!+\!\xi 
(\!\frac{3}{2}\!+\!\vec u^2\!)\!\Big )\!\hat C
\!+\!\xi u_iu_j \hat C^{ij}\!\!+\!\langle n|\hat E|n\rangle\!\right]
\ee
with
\be\label{cij}
C^{ij}\!\!=\!\!2\!\sum_r\langle r|p_i |n\rangle \langle n|p_j |r\rangle 
(E_r-E_n)\!\ln|\frac{E_*}{E_n-E_r}|
\ee
$\hat C\equiv\hat C^{ii}$ and
\ba\label{hate}
&\hat E&=4\pi Z\alpha\delta (\vec x)\left[\frac{19}{30}+\ln (\frac{m}{2E_*})\right.\nonumber\\
&+&\left.\xi\Big[-\frac{1}{30}-\frac{58}{45}\vec u^2+(\frac{3}{2}+\frac{2}{3}\vec u^2)
\ln(\frac{m}{2E_*})\Big]\right]\nonumber\\
&+&3\frac{Z\alpha}{r^3}\left[\frac{1}{4}+\xi\Big[\frac{1}{8}
-\frac{\vec u^2}{2}-(\vec u\cdot\nv)^2\Big]\right]\vec \sigma\cdot\vec L\\\nonumber
&-&\xi\frac{Z\alpha}{r^3} [3(\vec u\cdot\nv)^2-\vec u^2]\left[\frac{14}{15}
+2\ln (\frac{m}{2E_*})\right]\\\nonumber
&+&\frac{\xi}{2}\frac{Z\alpha}{r^2}\left[
\frac{7}{2}\vec u\cdot\nv\,\vec\sigma\cdot(\vec u\times\vec p)
-\vec\sigma\cdot(\vec u\times\nv)
\vec u\cdot\vec p\right]
\ea
where $E_*$ is a reference energy to be defined.
We have omitted operators with odd parity (such as 
$\vec u\times\nv\cdot\vec\sigma$) in (\ref{hate}), 
since their  expectation values vanish for states of definite parity. 

There is still an implicit dependence on $\xi$ and $\vec u$ 
coming from the Dirac states (due to the non coulombic behavior of the
electromagnetic source (\ref{po})), which modifies the answer in
(\ref{Elamb}) to
\be\label{efinal}
\!\!\delta E_n\!\!=\!\!\frac{\alpha}{3\pi m^2}\!\!\Big[\! \Big (\!1\!+\!\xi 
(\!\frac{3}{2}+\vec u^2\!)\!\Big )\!\hat C^0
\!\!+\!\xi u_iu_j \hat E^{ij}\!+\!^0\langle n|\hat E|n\rangle ^0\!\Big]
\ee
with 
\be\label{Eij}
u_iu_j \hat E^{ij}=u_iu_j \hat C^{ij}+\hat C'+(\,^0\langle n|\hat E_{\xi=0}|n\rangle'+\hbox{h.c.})
\ee
where $\hat C'$ groups all the terms in eq. (\ref{cij}) 
depending on  the perturbative corrections ($|n\rangle'$) and ($E_n'$)
to the states $|n\rangle$ and energies ($E_n$). These 
are needed not only for the $|n\rangle$ state
related to the level shift, but for all the intermediate states
introduced by (\ref{cij}) as well. 

If we now define the reference energy ($E_*$) as in the metric case \cite{IZ}, and evaluate 
eq. (\ref{efinal}) for the Lamb states, we can write the total 
contribution for, say,  the $2S_{1/2}$--$2P_{1/2}$  Lamb shift
(including the non radiative contribution (\ref{ED})) as: 
\vfil\eject
\be\label{EDQ}
\Delta E_L\!\!=\!\!\frac{m}{6\pi}(Z\al)^4\al\Big\{\!\!\ln\!{1\over\al^2}\!-\!2.084
\!+\!\xi\Big[\!\frac{3}{2}\!\ln\!{1\over\al^2}\!-\!4.534\ee
$$+ \vec u^2\!\!\left[\frac{\pi}{\al}
\!-\!3.486\!+\!\frac{2}{3}\ln\!{1\over\al^2}
\!-\! 0.011\,\cos^2\theta\right]\!\!+\! u_iu_j\Delta\hat\epsilon^{ij}\!\Big]\!\Big\}$$
 Here
$\theta$ represents the angle between the atom's quantization axis and the
frame velocity $\vec u$ and $\Delta\hat\epsilon^{ij}\equiv 
2\Delta\hat E^{ij}/((Z\al)^4m^3) $

Useful empirical information may be extracted by calculating the
gravitational 
redshift and time-dilation parameters associated with (\ref{EDQ}). 
In a redshift experiment the local energies at
emission $w_{em}$ and at reception $w_{rec}$ of a photon transmitted between
observers at different points in an external gravitational field are
compared in terms of
$Z=\frac{w_{em}-w_{rec}}{w_{em}} \equiv \Delta U\Big(1-\Xi\Big)$, whereas
experiments comparing
atomic energy transitions between the moving ($w_u$) and 
preferred ($w_0$) frames may be described via
$w_u=w_0(1-[A-1]\frac{\vec u ^2}{2})$.
The anomalous redshift ($\Xi$) and time dilation ($A$)
parameters may be computed from the anomalous passive and inertial
mass tensors using standard techniques \cite{will1,HW}.
After rescaling the action so that $\alpha \propto \sqrt{H/T}/\epsilon$,
we find for the Lamb shift transition (\ref{EDQ}) 
\be
\Xi^L=3.424\,\Gamma_0-1.318\,\Lambda_0
\label{lamlpi}
\ee
and
\be\label{al}   
\!\!\!1\!-\!A^L\!\!=\!\!\frac{\xi}{7.757}\!\!\left[\!\frac{\pi}{\al}\!+\!
3.074\!-\!0.011\cos^2\theta\!+\!\!
\frac{u_iu_j}{\vec u^2}\Delta\hat \epsilon^{ij}\!\right] 
\ee
with
\be\label{gamlam}
\Gamma_0\!\equiv\!\frac{T_0}{T_0'}\ln[\frac{T\epsilon^2}{H}]'\vert_0
\qquad
\Lambda_0\!\equiv\!\frac{T_0}{T_0'}\ln[\frac{T\mu^2}{H}]'\vert_0
\ee

We emphasize that qualitatively new information on the validity of 
the EEP will be obtained by  setting new empirical bounds on the
parameters $\xi$, $A_L$ and $\Xi_L$  which are associated with purely 
{\it leptonic} matter. Comparatively little is known about 
empirical limits on EEP-violation in this sector \cite{Hughes}.
Previous experiments have set the limits 
\cite{PLC} $|\xi_B| \equiv |1 -c_B^2|\,<\,6\,\times \,10^{-21}$ 
where $c_B$ is the ratio of the limiting speed of baryonic matter 
to the speed of  light, and \cite{redshift}
$|3\Gamma_{0}-\Lambda_{0} +2\Delta|<2\;\times\;10^{-4}$
and $\Gamma_{0}-\Gamma_{B0}\equiv
\Delta\equiv\Lambda_{0}- \Lambda_{B0}$ where $\Gamma_{B0}$ and 
$\Lambda_{B0}$ are quantities analogous to those in eq. 
(\ref{gamlam}) for baryons, where for simplicity we have assumed 
$T_{B0}=T_0$. This latter experiment involves interactions between 
nuclei and electrons and so does not (at least to the leading order to 
which we work) probe the leptonic sector in the manner 
that Lamb-shift experiments would.

The coefficient $A_L$ depends upon $\Delta\hat \epsilon^{ij}$, the 
evaluation of which
involves the numerical computation of the sum in (\ref{Eij}).
The contribution in eq.(\ref{al}) 
from the Dirac part of the energy (proportional to $\frac{1}{\al}$ ), 
produces an overall shift only.  Assuming that 
EEP-violating contributions to $\Delta E_L$ are
bounded by the current level of precision for the Lamb shift 
\cite{Eides}, and that
$|\vec{u}| < 10^{-3}$ \cite{will1}, we find the dominant contribution to
(\ref{al})  is due to the purely radiative part of (\ref{EDQ}), yielding
the bound $|\xi| < 10^{-5}$. This limit is comparable to that 
noted in a different context by Greene {\it et. al.} \cite{Greene}.
We note that anisotropic effects in (\ref{al}) arise solely 
from radiative corrections.

Improvement on such bounds will be a challenge to experimentalists
because of the intrinsic uncertainties of excited states of Hydrogenic
atoms. Setting empirical bounds on $\Xi_L$ will involve measuring the 
frequency shifts of an atomic clock based on the Lamb shift transition, 
either by comparing two such transitions at different points in a 
gravitational potential or by performing a 
`clock-comparison' type of experiment between a `Lamb-shift clock' and 
some other atomic frequency standard \cite{will1}. 
The former experiment would appear unfeasible 
since the anticipated redshift in the background potential of
the earth ($\approx 10^{-9}$) is
much smaller than any foreseeable improvement
in the precision of Lamb-shift transition measurements \cite{Eides}. 
One would at least need to  perform the experiment in a
stronger gravitational field (such as on a satellite in close solar 
orbit) with 1-2 orders-of-magnitude improvement in precision.
The latter measurement is, in principle, sensitive to the 
absolute value of the local potential \cite{Hughes,Good}, whose 
magnitude has recently been estimated to be $\approx 10^{-5}$ due to the
local supercluster \cite{Kenyon}. The best hope would appear to be 
in exploiting anticipated improvements in precision \cite{Eides} to
obtain better bounds on $\xi$ via (\ref{EDQ}).

Finally, we note that our calculation has assumed that positrons and
electrons have equivalent couplings to the gravitational field. If
we drop this assumption \cite{Schiff}, we find that there is an 
additional contribution to (\ref{EDQ}) due to $\xi_{e^+} 
\neq \xi_{e^-}$. This contribution is due entirely to radiative 
corrections.  Making the same comparisons as above, we find the most 
stringent bound on this quantity 
to be $|\xi_{e^+} | < 10^{-3}$ from present experiments.  In a similar
vein, one could also consider tests of the EEP for muonic atoms to
obtain bounds on $\xi_{\mu}$.  We shall report on this elsewhere
\cite{catlamb}.

The intrinsically quantum-mechanical character 
of the radiative corrections will, we hope,  motivate
the development of new EEP experiments based 
on the Lamb shift transition, thereby extending
our understanding of the validity of the
equivalence principle into the regime of quantum-field theory.

\section*{Acknowledgments}
This work was supported in part 
by the Natural Sciences and Engineering Research
Council of Canada.  
We are grateful to C.M. Will for his initial encouragement in
this work and to E. Hessels for discussions.

\end{document}